\documentclass[11pt,a4paper]{article}
\usepackage{latexsym}
\newtheorem{theorem}{Theorem}
\newtheorem{lemma}{Lemma}
\newtheorem{corollary}{Corollary}
\newtheorem{fact}[lemma]{Fact}
\newtheorem{definition}{Definition}
\newcommand{\qed}{\hfill$\Box$}

\newcommand{\N}{{\rm I\!N}}
\newcommand{\tensor}{\otimes}
\newcommand{\trace}{{\rm Tr}}
\newcommand{\prob}{{\rm Prob}}
\newcommand{\size}[1]{\left|#1\right|}

\newcommand{\ket}[1]{\left|#1\right\rangle}
\newcommand{\bra}[1]{\left\langle #1\right|}
\newcommand{\braket}[2]{\left\langle #1\!\mid\! #2\right\rangle}
\newcommand{\ketbra}[2]{\ket{#1}\!\bra{#2}}
\newcommand{\norm}[1]{\left\|\,#1\,\right\|}
\newcommand{\trnorm}[1]{\norm{#1}_{\rm t}}
\newcommand{\set}[1]{{\left\{#1\right\}}}

\newcommand{\co}{{\cal O}}
\newcommand{\ch}{{\cal H}}

\newcommand{\aitch}{{\mathcal H}}
\newcommand{\kay}{{\mathcal K}}

\newcommand{\brho}{{\mathbf \rho}}

\newcommand{\trn}[1]{\trnorm{#1}}

\newcommand{\density}[1]{\ketbra{#1}{#1}}

\newcommand{\X}{{\mathcal X}}
\newcommand{\Y}{{\mathcal Y}}
\newcommand{\cp}{{\mathcal P}}

\begin{document}

\title{\Large\bf On quantum and approximate privacy}

\author{Hartmut Klauck
\thanks{Supported by NSF grant CCR-9987845. Work mostly done while at CWI and supported by
the EU 5th framework program QAIP IST-1999-11234 and by NWO grant 612.055.001.}\\
School of Mathematics\\
     Institute for Advanced Study\\
     Princeton, NJ08540, USA\\
     {\tt klauck@ias.edu}}

\date{}
\maketitle

\begin{abstract}
This paper studies privacy and secure function evaluation in communication complexity. The focus is on quantum versions of the model and on
protocols with only approximate privacy against honest players. We show that the privacy loss (the minimum divulged information) in computing a
function can be decreased exponentially by using quantum protocols, while the class of privately computable functions (i.e., those with privacy
loss 0) is not enlarged by quantum protocols. Quantum communication combined with small information leakage on the other hand makes certain
functions computable (almost) privately which are not computable using either quantum communication without leakage or classical communication
with leakage. We also give an example of an exponential reduction of the communication complexity of a function by allowing a privacy loss of
$o(1)$ instead of privacy loss 0.
\end{abstract}

\section{Introduction}
\label{sec-intro}

{\it Mafiosi} Al and Bob, both honest men, claim rights to protect a
subset of the citizens of their hometown. To find out about possible
collisions of interest they decide to communicate and find out whether
there is a citizen they both intend to protect. Of course they would like to
do this in a way that gives each other as little information as possible
on the subset they think of. In other words, they want to compute a
function with as much privacy as possible, rather than caring about
the communication cost inclined. This problem is one of the kind
studied in the theory of private computation resp.~secure function evaluation,
initiated by Yao \cite{Y82}. Another
example is the two millionaires' problem, in which Al and Bob try to determine who is
richer, but without revealing more about their actual wealth.

Informally a protocol for the computation of some function on inputs distributed to several players
is {\it private}, if all information that can
be deduced by one player during a run of the protocol can also be deduced from the player's input and the function
value alone. A function is
private, if it can be computed by a private protocol. Generalizing this a function is said to have {\it privacy loss} $k$,
if the minimum information divulged to the other players is $k$ in any protocol computing the function. In this definition we
use an information theoretic measure for the privacy loss. Alternatively the information {\it leakage} of a protocol may
be measured as a distance between message states that must be ``almost indistinguishable'' for a player. This setup
generalizes several cryptographic scenarios, see \cite{Y82}.

There are some variants and twists to this model. One can
distinguish computationally secure and information theoretically
secure protocols. The first variant is studied e.g.~in \cite{Y82} and
\cite{GMW87}. Multiparty protocols in the information theoretically secure
setting are given in \cite{BGW88} and \cite{CCD88}. A second kind of
variations concerns the type of players. Basically ``honest but
curious'' and ``malicious'' (or ``byzantine'') players have been
considered in the literature. The first type of players sticks to the protocol but tries
to get information by running some extra program on the messages
received. The second type of players deviates arbitrarily
from the protocol to get as much information as possible. Furthermore
protocols may be deterministic, randomized, or use the possibilities
offered by quantum communication. A quantum variant of
(information-theoretically) secure
multiparty computation with malicious players has been
investigated recently in \cite{CGS02}. A thorough study of secure
quantum computation with honest players seems to be missing,
especially in the two-player case. Lo has investigated the case of
one-sided secure quantum computation \cite{L97}, in which only one
player learns the function value. Certain aspects of general two-party
secure quantum computation are discussed in \cite{LC98}.

To compare the possible combinations of the above choices concerning the
underlying model consider the following facts known in the non-quantum
setting. The two millionaires' problem
has a computational solution \cite{Y82} relying on the existence of
one-way functions, but it cannot be solved in the
information theoretic sense \cite{CK91} (not even among honest players), i.e., some
cryptographic hardness assumption has to be used. A variant of the
millionaires' problem that can actually be solved with information theoretic privacy
for honest players is the identified minimum problem, in which the wealth of the less
rich player and his identity is revealed, but no additional information about the
wealth of the other player \cite{K92}. Secure function evaluation among two
dishonest players without computational restrictions is
usually impossible. Information-theoretically secure multiparty
protocols ($\ge 3$ players) with more than two thirds of all
players being honest are possible for all functions \cite{BGW88}, in
the computationally secure setting it is possible to compute all
functions when more than one half of all players are honest \cite{GMW87}.

In this paper we concentrate on information theoretical security and honest
players. While considering only honest players seems to strongly
restrict the model, it is important
for several reasons. First, understanding honest players is a
prerequisite to understanding actively cheating players. Secondly, these
players capture ``passive'' attacks that cannot be detected in any way, which
might be an important motivation for curious players to follow such a
strategy. Furthermore \cite{GMW87} gives a quite general reduction
from multiparty protocols with honest majority to protocols with only honest
players (in the computationally secure setting). Other motivations for
considering this model include close connections to complexity
measures like circuit size \cite{KOR99}. Privacy loss may also be viewed as a
complexity measure, having some useful connections to communication
complexity exploited e.g.~in \cite{CSWY01,BJKS02}.

We focus on the following aspects of private
computing. Al and Bob have heard that quantum computers can break
cryptographic schemes classically assumed to be secure, so they do not want to rely
on computational solutions\footnote{Actually it is quite possible that
quantum one-way permutations exist, see e.g.~\cite{BBBV97}.}.
They are interested in whether quantum
communication enlarges the set of privately computable functions
or substantially decreases the privacy loss of functions. Furthermore
they  are interested in whether it is possible to decrease the communication
cost of a protocol by allowing leakage of a small amount of
information. We concentrate on the two player model in this paper,
though some of the results have implications for the multiparty
setting, which we mention in the conclusions.

The functions we mainly consider in this paper are the disjointness
problem $DISJ_n$, in which Al and Bob each receive a subset of a size
$n$ universe, and have to decide whether their subsets are disjoint or
not, and the identified minimum problem $IdMin_n$, in which Al and Bob
receive numbers $x,y$ from 0 to $2^n-1$, and the output is $2x+1$,
if $x\le y$, and $2y$ otherwise.

The type of players we investigate are honest but
curious. This means they stick to the protocol, but otherwise try
anything they can to get information about the other player's input. In
the quantum case a major point will also be whether the players might
be trusted to not quit the protocol before the end. Our main model
will measure the maximum information obtainable over all the rounds,
not only the
information obtainable at the end of the protocol. This corresponds to players
that might quit the protocol before its end\footnote{Al has the habit
  of shooting his guests after dessert, which may be well before the
  end of the protocol.}.
The other model of nonpreemptive players will be
investigated also, but here every function turns out to be computable almost
privately and at the same time efficiently in the quantum case.

Our main results are the following. We show that the quantum protocol for disjointness with communication complexity $O(\sqrt{n}\log n)$ given in
\cite{BCW98} can be adapted to have privacy loss $O(\log^2 n)$. We proceed to show that any classical bounded error protocol for disjointness
divulges $\Omega(\sqrt{n}/\log n)$ bits of information. Thus Al and Bob are highly motivated to use the quantum protocol for privacy reasons.
Note also that any (even nonprivate) classical randomized protocol for disjointness needs communication $\Omega(n)$ \cite{KS92}. Every
(nonprivate) quantum protocol for disjointness needs communication $\Omega(\sqrt{n})$, as recently established in \cite{R03}.

We then show that the class of privately computable functions is not
enlarged by using quantum computers, i.e., every function that can be
computed privately using a quantum protocol can also be computed privately by a deterministic
protocol. This result leads to the same characterization
of privately computable functions as in the classical case.
We furthermore show that allowing a small leakage combined with quantum communication allows to
compute  Boolean functions which are nonprivate.
This does not hold for both quantum communication
without leakage and classical communication with leakage. We also
analyze a tradeoff between the number of communication rounds and the leakage
required to quantum compute a nonprivate function.

We then turn to the question, whether leakage can decrease
the communication complexity and show that $IdMin_n$
can be computed with leakage $1/poly(n)$ and communication
$poly(n)$, while any perfectly private (quantum) protocol needs
$\Omega(2^n)$ rounds and communication. Thus a tiny leakage
reduces the communication cost exponentially. It has been known previously
\cite{BCKO93} that one bit of privacy loss in the ``hint sense'' can decrease the
communication complexity exponentially, but in our result the
privacy loss is much smaller, and the function we consider is more
natural than the example in \cite{BCKO93}.

The paper is organized as follows. In the next section we give the
necessary definitions and some technical results. Section 3 describes
the result about an exponential decrease of privacy loss by using
quantum communication. Section 4 discusses the set of functions
computable by private or almost private quantum protocols. Section 5
shows how allowing very small privacy loss can decrease communication
complexity. In section 6 we give conclusions and some open problems.

\section{Preliminaries}
\label{sec-pre}

In this section we first describe the communication model we study,
then the (quantum) information theoretic  notions used, and finally
discuss privacy definitions. For introduction to quantum computing see
e.g.~\cite{NC00}.

\subsection{The communication complexity model}

In the quantum communication complexity model~\cite{Y93}, two parties
Al and Bob hold qubits. When the game starts Al holds a
superposition~$\ket{x}$ and Bob holds~$\ket{y}$ (representing the input
to the two players), and so the
initial joint state is simply~$\ket{x} \tensor \ket{y}$. Furthermore
each player has an arbitrarily large supply of private qubits in some
fixed basis state.
The two parties then play in {\it rounds}. Suppose it is Al's turn to play.
Al can do an arbitrary unitary transformation on his qubits
and then send one or more qubits to Bob.
Sending qubits does not change the overall superposition, but rather
changes the ownership of the qubits, allowing Bob to
apply his next unitary transformation on the newly received qubits.
Al may also (partially) measure his qubits during his turn.
At the end of the protocol, one player makes a measurement and sends
the result of the protocol to the other player.
The overall number of message exchanges is called the number of rounds.
In a {\it classical} probabilistic protocol the players may only exchange
messages composed of classical bits.

The complexity of a quantum (or classical) protocol is the number of qubits
(respectively, bits) exchanged between the two players in the worst case.
We say a protocol {\em computes\/}
a function~$f : \X \times \Y \mapsto \cal{Z}$
with~$\epsilon \ge 0$ error if, for any input~$x \in \X,y \in \Y$,
the probability
that the two players compute~$f(x,y)$ is at least $1-\epsilon$.

Sometimes we want to relax the above correctness requirement.
We say a protocol~$\cp$ computes $f$ with~$\epsilon$ error
with respect to a distribution~$\mu$ on~$\X \times \Y$,
if
$$\prob_{(x,y) \in \mu, \cp}( \cp(x,y)=f(x,y)) \;\;\ge\;\; 1-\epsilon.$$

A randomized classical or a quantum protocol has access to a {\it
  public coin}, if the
players can flip a {\em classical} coin and both read the result without
  communication. If not mentioned otherwise we do not consider this
  variant of the model.

The {\em communication matrix} of a function $f(x,y)$ is the matrix with
rows labelled by the $x$'s, columns labelled by the $y$'s, and
containing $f(x,y)$ at position $x,y$.

A {\em rectangle} in the communication matrix is a submatrix indexed by
the product of a subset of the row-labels and a subset of the
column-labels. A rectangle is {\it monochromatic}, if all its entries
are the same.

\subsection{Information theory background}
\label{sec-info-theory}

The quantum mechanical analogue of a random variable is
a probability distribution over superpositions, also
called a {\em mixed state}. For the mixed state~$X = \{p_i,\ket{\phi_i}\}$, where
$\ket{\phi_i}$ has probability~$p_i$, the {\em density matrix\/}
is defined as~$\brho_X  = \sum_i p_i \ketbra{\phi_i}{\phi_i}$.
Density matrices are Hermitian, positive semidefinite, and have trace
$1$. I.e., a density matrix has only real
eigenvalues between zero and one, and they sum up
to one.

The {\em trace norm\/} of a matrix~$A$ is defined as~$\trn{A}
= \trace\,{\sqrt{A^\dagger A}}$, which is equal to the sum of the
magnitudes of the singular values of~$A$.
Note that if $\rho$ is a density matrix, then it has trace norm one.
If $\phi_1,\phi_2$ are pure states then:
\begin{equation}\trnorm{\density{\phi_1} - \density{\phi_2}}
    ~~=~~ 2 \sqrt{ 1 - \size{\braket{\phi_1}{\phi_2}}^2}.\end{equation}

 The following important fact (the Kraus representation theorem)
 characterizes the physically allowed quantum operations on density
 matrices (trace preserving completely positive superoperators)
 in terms of adding blank qubits, doing unitary transformations, and tracing out (see
 \cite{NC00}). Hence we can restrict our considerations to these operations.

 \begin{fact}\label{fact:Kraus}
   The following statements are equivalent:
 \begin{enumerate}
 \item An operation $T$ sending density matrices over $H_1$ to
   density matrices over $H_2$ is physically allowed (i.e., trace
   preserving and completely positive).
 \item There is a Hilbert space $H_3$ with $dim(H_3)\le dim(H_1)$ and
   a unitary transformation $U$, such that for all density matrices
   $\rho$ over $H_1$ the result of $T$ applied to $\rho$ is
  \[trace_{H_1\otimes H_3}[U(\rho\otimes|0_{H_3\otimes
    H_2}\rangle\langle0_{H_3\otimes H_2}|)U^\dagger ].\]
  \end{enumerate}
  \end{fact}

  So allowed operations can be simulated by
  adding some blank qubits, applying a unitary
  transformation and ``dropping'' some qubits.

Another useful theorem states that for two mixed states $\rho_1,\rho_2$
their distinguishability is reflected in $\trn{\rho_1-\rho_2}$
\cite{AKN98}:

\begin{fact}
\label{fact:trace}
Let~$\rho_1,\rho_2$ be two density matrices on the same space~$\ch$.
Then for any measurement~$\co$,
$$
\norm{\rho_1^\co - \rho_2^\co}_1 ~~\le~~ \trn{\rho_1-\rho_2},
$$
where~$\rho^\co$ denotes the classical distribution on outcomes
resulting from the measurement of~$\rho$, and~$\norm{\cdot}_1$
is the~$\ell_1$ norm. Furthermore, there is a measurement $\co$, for which
the above is an equality.
\end{fact}

The {\em Shannon entropy\/}~$H(X)$ of a classical random variable~$X$
and {\em mutual information\/}~$I(X:Y)$ of a pair of random variables~$X,Y$
are defined as usual (see e.g.~\cite{CT91}).

The {\em von Neumann entropy\/}~$S(\brho)$ of a density
matrix~$\brho$ is defined as~$S(\brho) = - \trace\,\brho\log\brho =
- \sum_i \lambda_i \log \lambda_i$, where~$\{\lambda_i\}$ is the
multi-set of all the eigenvalues of~$\brho$. Notice that the eigenvalues
of a density matrix form a probability distribution.
For properties of this function see \cite{NC00}.

We use the following fact about the continuity of entropy (see theorem
16.3.2 in \cite{CT91} and theorem
11.6 in \cite{NC00}).
\begin{fact}\label{fact:contin}
Let $p,q$ be distributions on $\{0,1\}^n$ with $d=||p-q||_1=\sum_x |p_x-q_x|\le
1/2$. Then \[|H(p)-H(q)|\le d\cdot n-d\log d.\]

Let $\rho,\sigma$ be states in a $2^n$ dimensional Hilbert space with
$d=\trn{\rho-\sigma}\le 1/e$.
Then  \[|S(\rho)-S(\sigma)|\le d\cdot n-d\log d.\]
\end{fact}

An immediate corollary of Jensen's inequality is the following:

\begin{fact}\label{fact:maxent}
For $x\in\{0,1\}^n$ let $0\le p_x$ and $\sum_x p_x\le\gamma\le 1$.
Then \[-\sum_x p_x\log p_x\le\gamma n-\gamma\log \gamma.\]
\end{fact}

For a bipartite quantum state $\rho_{XY}$ we define the ``mutual information''~$I(X:Y)$
as $I(X:Y)=S(X)+S(Y)-S(XY)=S(\rho_X)+S(\rho_Y)-S(\rho_{XY})$, where~$\rho_X,\rho_Y$ are the reduced density
matrices on the systems $X,Y$.
We also define  {\em conditional\/} mutual information~$I(X:Y|Z)$ as follows:
$$
I(X:Y|Z) ~~=~~ S(XZ) + S(YZ) - S(Z) - S(XYZ).
$$

We will employ the following facts from \cite{KNTZ01}.
\begin{fact}[Average encoding theorem]
\label{fact:average}
Let $x \mapsto \rho_x$ be a quantum encoding
mapping~an $m$ bit string~$x\in \set{0,1}^m$ into a mixed
state with density matrix $\rho_x$.
Let~$X$ be distributed over~$\set{0,1}^m$, where $x\in\set{0,1}^m$ has probability
$p_x$, let~$Q$ be the
register holding the encoding of~$X$ according to this map, and
let $\bar{\rho} = \sum_x p_x\rho_x$.
Then,
\begin{eqnarray*}
\sum_x p_x\trn{\bar{\rho} - \rho_x}
    & \le & \sqrt{(2 \ln 2)\, I(Q:X) }.
\end{eqnarray*}
\end{fact}

    A {\it purification} of a mixed state with
    density matrix $\rho$ over some Hilbert space
    $\cal H$ is any pure state $|\phi\rangle$ over
    some space ${\cal H}\otimes {\cal K}$ such that $trace_{\cal K}
    |\phi\rangle\langle\phi|=\rho$.

\begin{fact}[Local transition theorem]
\label{fact:local}
Let~$\brho_1,\brho_2$ be two mixed states with support in a  Hilbert
space $\aitch$,
$\kay$ any Hilbert space of dimension at least~$\dim(\aitch)$, and
$\ket{\phi_i}$ any purifications of the~$\brho_i$ in $\aitch \tensor
\kay$.
Then, there is a local unitary transformation~$U$ on~$\kay$ that
maps~$\ket{\phi_2}$ to~$\ket{\phi'_2} = I\tensor U\ket{\phi_2}$ such
that
$$\trn{ \ketbra{\phi_1}{\phi_1} - \ketbra{\phi'_2}{\phi'_2}}
\;\;\le\;\; 2 \trn{\brho_1-\brho_2}^{\frac{1}{2}}.$$
\end{fact}

\subsection{Privacy}

Given a protocol a player is {\em honest} if for each input and all messages
he receives he sends exactly
the messages prescribed by the protocol.  All operations
are allowed as long as this requirement is met. It is e.g.~allowed
to copy a classical message to some additional storage (if it is known
that the message is classical). Copying
general unknown quantum states is, however, impossible \cite{WZ82}.

We state this requirement a bit more formal in the following way. In a
quantum protocol as defined above the actions of the players are
defined as a series of unitary transformations plus the sending of a certain
choice of qubits. For player Al to be honest we demand that for all rounds $t$
of the protocol, for all inputs $x$, and for all sequences of pure
state messages he may have received in the previous rounds, the density
matrix of the message in the next round equals the density matrix
defined by the protocol and the input. Note that in a run of the
protocol the player might actually receive mixed state messages, but
the behavior of the player on these is defined by his behavior on
pure state messages.

We define the privacy loss of a protocol as follows. Let
$\rho_{ABXY}$ denote a state of the registers containing Al's private
qubits in $A$, Bob's private qubits in $B$, Al's input in $X$, Bob's
input in $Y$. We assume that the (classical) inputs are never erased
by the players.

For a distribution $\mu$ on the inputs to a protocol computing $f$ the information
divulged to Bob at time $t$ is $L(t,B,\mu)=I(B:X|Y,f(X,Y))$, for the state
$\rho^{(t)}_{ABXY}$ of the protocol at time $t$ induced by the
distribution $\mu$ on the inputs. Symmetrically we define Al's loss
$L(t,A,\mu)$. The privacy loss of a protocol is the supremum
of $L(t,\cdot,\mu)$ over all $t$ and $A,B$ and all $\mu$.

The {\em privacy loss} $L_\epsilon(f)$ of a function $f$ is the infimum privacy loss
over all quantum protocols computing $f$ with error $\epsilon$.
The classical privacy loss $CL_\epsilon(f)$ is defined analogously,
with the infimum over all classical randomized protocols.

A function $f$ is said to be {\em private}, if $CL_0(f)=0$. It is known that
$CL_\epsilon(f)=0$ with $\epsilon<1/2$ holds only for private
functions \cite{K92}.

Note that in the above definition we have assumed that the information
available to a player is small in all rounds. Thus even if one player
decides to quit the protocol at some point the privacy loss is
guaranteed.

If we consider only the final state of the protocol in our definition
we call the players {\em honest and nonpreemptive}. For a classical protocol
there is no difference between these two possibilities, since the
information available only increases with time. In the quantum
case, however, this is not true.

The information divulged by a nonprivate protocol can also be measured in
a different way, namely via distinguishability, see \cite{CK91}. Let
$\rho_{AB}^{xy}$ denote the state of Al's and Bob's qubits in some
round for inputs $x,y$, and let $\rho^{xy}_A$ resp.~$\rho_B^{xy}$
denote the reduced density matrices on Al's and Bob's qubits.
A protocol is said to {\it leak} at most $\delta$ to Bob,
if for all $x,x'$ and $y$ with $f(x,y)=f(x',y)$ it is true that
\[\trn{\rho_{B}^{xy}-\rho_{B}^{x'y}}\le\delta.\] This means that no
quantum operation on Bob's qubits can distinguish the two states better
than $\delta$. Thus there is a limit on Bob's ability to distinguish
Al's inputs as long as changing these does not change the output.
An analogous definition is made for Al. We say a protocol leaks at
most $\delta$, if the maximum leakage to a player in any round is at
most $\delta$.

Two more definitions of (classical) privacy loss are considered in
\cite{BCKO93}. In the first variant there is not one protocol which is
good against all distributions on the inputs, but for each distribution there may be
one specialized protocol, for which the privacy loss is measured. The
second definition is privacy loss in the {\it hint sense}, here a function
$h$ is computed privately instead of a function $f$, and $f(x,y)$ can
be computed from $h(x,y)$. The privacy
loss is the difference between the logs of the ranges of $f$ and $h$.
We generally use privacy loss with respect to the definition we have given first.

Let us show that our standard definition of privacy loss and the first
alternative definition mentioned above are asymptotically equivalent for
randomized and quantum protocols. The following lemma is a
consequence of the standard Yao principle (von Neumann duality), see \cite{KN97}.

\begin{lemma}\label{lem:yao}
The following statements are equivalent in the sense that if one is true for some values $\epsilon,\delta$, then
the other is true with values $2\epsilon,2\delta$.
\begin{itemize}
\item There is a randomized [quantum] public coin protocol for a function $f$ with
communication $c$, error $\epsilon$, privacy loss
$\delta$ against all distributions $\mu$ on inputs.
\item
For every distribution $\mu$ on inputs there is a deterministic
[quantum] protocol for $f$ with communication $c$, error $\epsilon$, and
privacy loss $\delta$ on that distribution.
\end{itemize}\end{lemma}

{\sc Proof:}
The direction from randomized [quantum] to distributional deterministic [quantum] protocols follows by observing that
a public coin protocol is really a probability distribution on deterministic [quantum] protocols, and for each distribution
$\mu$ on the inputs the expected error (when picking a deterministic [quantum] protocol) is $\epsilon$, and likewise the expected
privacy loss is $\delta$. Now due to the Markov inequality for each $\mu$
there must be one deterministic [quantum] protocol that has error at most $2\epsilon$ and privacy loss at most $2\delta$
simultaneously.

For the other direction assume the second statement holds, then combine error and privacy loss into one parameter by setting
$para(P,\mu)=err(P,\mu)\cdot \delta+loss(P,\mu)\cdot\epsilon$, where $err(P,\mu)$ denotes the error of a deterministic [quantum]
protocol $P$ on $\mu$, and $loss(P,\mu)$ the privacy loss. Note that we are guaranteed that for each $\mu$ there is a $P$
with $para(P,\mu)\le2\epsilon\delta$.
Now the standard Yao principle gives us a single public coin randomized [quantum] protocol
that has expected $para(P,\mu)\le2\epsilon\delta$ for all $\mu$.
Such a protocol must have expected error at most $2\epsilon$ and expected privacy loss at most $2\delta$ for all $\mu$.
\qed

Our definition of communication complexity allows no public coins,
however.
If we are only interested in the privacy loss, one of the players may simply
flip enough coins and communicate them to the other player, then they
simulate the public coin protocol. This increases the communication,
but none of the other parameters.

We need another result to get rid of the public coin at a lower cost
in randomized protocols,
if the leakage resp.~privacy loss is very small. First consider the
following lemma concerning leakage, proved completely analogous to the
results in \cite{Ne91}.

\begin{lemma}\label{lem:Newm}
Let $f:\{0,1\}^n\times\{0,1\}^n\to\N$ be computable by a
randomized [quantum] protocol with error $\epsilon$, that uses public
classical randomness
and $c$ bits of communication and leaks $\delta$.

Then for all $\gamma>0$ there is a randomized [quantum]
protocol for $f$ with
error $(1+\gamma)\epsilon$,
leakage $(1+\gamma)\delta$, and communication
$O(c+\log n+\log(1/\delta)+\log(1/\gamma)+\log(1/\epsilon))$ that uses
no public coin.
\end{lemma}

If leakage is small a bound on privacy loss is implicit.

\begin{lemma}\label{lem:leakpriv}
 Let $f:\{0,1\}^n\times\{0,1\}^n\to\N$ be computable by a
randomized [quantum] protocol (with error
but using no public randomness) that has leakage $\delta\le 1/e$.

Then the same protocol has privacy loss at most
$n\cdot\delta-\delta\log\delta$.
\end{lemma}

{\sc Proof:} First consider the case of classical protocols. Let
$\mu$ be any distribution on the inputs. The
distribution on the values of
Bob's and Al's private storage $A,B$, when the inputs $X,Y$ are drawn according to $\mu$
is denoted $\rho_{ABXY}$ (at some point in the protocol). If inputs are fixed to $x,y$ the resulting
(normalized) distribution is denoted $\rho_{AB}^{xy}$.
The leakage requirement states that for $x,y,y'$
with $f(x,y)=f(x,y')$ we have $||\rho_{A}^{xy}-\rho_{A}^{xy'}||_1\le\delta$.
For $x,y$ with $f(x,y)=z$ let
\[\rho_A^{x}(z)=\sum_{y':f(x,y')=z}
\frac{\mu(x,y')}{\sum_{a:f(x,a)=z } \mu(x,a)} \rho_A^{xy'}.\]
Then due to convexity for all $x,y$ \[||\rho_A^{xy}-\rho_A^x(f(x,y ))||_1\le\delta.\]
The continuity of entropy (fact \ref{fact:contin}) and taking the expectation over $x,y$ acording to $\mu$ then gives us
\[E_{x,y} [H(\rho_A^x(f(x,y)))-H(\rho_A^{xy})]\le
  n\delta-\delta\log\delta.\]
The left hand side equals
\[H(A|X,f(X,Y))-H(A|X,Y,f(X,Y))=I(A:Y|X,f(X,Y)).\]
The leakage to Bob is analyzed in the same way.
The quantum case is completely analogous.
\qed

Now we can say something about privacy loss and private coins.

\begin{lemma}\label{lem:privlcoins}
Let $f:\{0,1\}^n\times\{0,1\}^n\to\cal Z$ be computable by a
randomized [quantum] protocol with error
$\epsilon$ (using a public classical coin) and $c$ bits of
communication, that has privacy loss $\delta\ge 1/2^{2n}$.

Then there is a randomized [quantum] protocol for $f$ with error $2\epsilon$,
privacy loss $O(n\cdot\sqrt{\delta})$, and communication
$O(c+\log n+\log(1/\delta)+\log(1/\epsilon))$ that uses no public coin.
\end{lemma}

{\sc Proof:} Consider the case of classical protocols.
Given the protocol, for, say, player Al, all distributions and all
rounds we have $I(A:Y|X,f(X,Y))\le\delta$. Denote the distribution of
the values of Al's register
$\rho_{A}^{xy}$ for some inputs $xy$, let $\rho_{A}^x(z)$ denote the
distribution in
which $x$ and $z=f(x,y)$ are fixed, but $y$ is random.
Then
\begin{eqnarray*}
&&E_zE_{x,y:f(x,y)=z} ||\rho^{xy}_A-\rho^{x}_A(z)||_1\\
&=& E_z E_{x|z} E_{y|zx} ||\rho^{xy}_A-\rho^x_A(z)||_1\\
&\stackrel{*}{\le}& E_{z,x} \sqrt{2\ln(2)I(A:Y|X=x,f(X,Y)=z)}\\
&\le& \sqrt{2\ln(2) E_{z,x}I(A:Y|X=x,f(X,Y)=z)}\mbox{ with Jensen's inequality}\\
&\le&\sqrt{2\ln(2)\delta}\end{eqnarray*} with (*) due to fact
\ref{fact:average}, where for a fixed $z,x$ the $y$ are coded (on the
induced distribution) as $\rho_A^{xy}$ and the average code is
$\rho_A^x(z)$.
 Hence \[E_{x,y,y':f(x,y)=f(x,y')} ||\rho^{xy}_A-\rho^{xy'}_A||_1\le
2 \sqrt{2\ln(2)\delta}.\] Since this holds for all distributions, the same
holds for all $x,y,y'$ with $f(x,y)=f(x,y')$, thus the protocol leaks at most
$2\sqrt{2\ln(2)\cdot \delta}$.

Invoking lemma \ref{lem:Newm} with $\gamma=1$ we get a
protocol with the desired communication complexity and error, and leakage
$O(\sqrt{\delta})$ using no public coin.
Then an application of the previous lemma completes the proof for leakage at most $1/e$. For larger leakage the lemma is trivial.

The quantum case is completely analogous.
\qed

Finally, the following lemma states that independent repetitions of a
randomized protocol allow to decrease the error probability
with a reasonable increase in privacy loss.

\begin{lemma}\label{lem:boost}
Let $P$ be any randomized protocol for a function $f$ with error
$1/3$, privacy loss $l$, and communication $c$.

Then there is a randomized protocol for $f$ with error $1/2^k$,
privacy loss $O(k\cdot l)$, and communication $O(k\cdot c)$.
\end{lemma}

{\sc Proof:}
Repeat the protocol $t=O(k)$ times independently. By standard
considerations taking the majority output yields the desired error
bound and increases the
communication as desired. We now show that the privacy loss is also as stated.

To see this consider the global state $\rho_{ABXY}$, where Al's
storage $A$ consists of $A_1,\ldots,A_t$ for the $t$ repetitions. W.l.o.g.~$A_i$
contains a message history of the $i$th repetition of the
protocol. Note that $I(A_i:A_j|X=x,Y=y)=0$ for all $i\neq j$ and all $x,y$, if Al plays
honest, since he is forced to send messages as in a completely new run
of the protocol for all histories of the first $i$ repetitions, i.e.,
using fresh randomness. Then
\begin{eqnarray*}
&&I(A_1,\ldots,A_t:Y|X=x,f(X,Y)=z)\\
&=&H(A_1,\ldots,A_t|X=x,f(X,Y)=z)\\
&-&H(A_1,\ldots,A_t|Y,X=x,f(X,Y)=z)\\
&=&H(A_1,\ldots,A_t|X=x,f(X,Y)=z)\\
&-&E_y H(A_1,\ldots,A_t|Y=y,X=x)
\end{eqnarray*}
for all $x,z$, with the expectation over $y\in Y$ under the
distribution conditioned on $f(x,y)=z$.
And due to the subadditivity of entropy this is at most \[\sum_i H(A_i|X=x,f(X,Y)=z)-E_y
H(A_1,\ldots,A_t|Y=y,X=x).\] The latter term equals \[E_y\sum_i
H(A_i|Y=y,X=x),\] since for a fixed input $x,y$ the random
variables $A_i$
are independent. So we get
\begin{eqnarray*}
&&\sum_i H(A_i|X=x,f(X,Y)=z)\\
&-&\sum_i E_y H(A_i|Y=y,X=x)\\
&=&\sum_i I(A_i:Y|X=x,f(X,Y)=z).
\end{eqnarray*}
Hence, with $I(A_i:Y|X,f(X,Y))=E_{x,z} I(A_i:Y|X=x,f(X,Y)=z)$:
\begin{eqnarray*}
&&I(A_1,\ldots,A_t:Y|X,f(X,Y))\\
&\le& \sum_i I(A_i:Y|X,f(X,Y))\\
&=& t\cdot I(A_1:Y|X,f(X,Y)).\hspace{7.5cm}\Box\end{eqnarray*}

\section{An exponential decrease in privacy loss}

In this section we give an example of a function that can be computed
with an exponentially smaller privacy loss in the quantum case than in
the classical case. This function is the disjointness problem, and the
quantum protocol we consider is the protocol due to Buhrman, Cleve,
and Wigderson given in \cite{BCW98}. In fact we describe a general way to
protect a certain type of protocols
against large privacy loss.

We now roughly sketch how the protocol works, and then how to make it
secure. The protocol is based on a general simulation of black-box
algorithms given in \cite{BCW98}. A black-box algorithm for a function
$g$ is turned into a communication protocol for a function $g(x\wedge
y)$ for the bitwise defined operation $\wedge$. The black-box
algorithm for OR is the famous search algorithm by Grover \cite{G96}
or rather its variant in \cite{BBHT96}. The important feature of the
protocol for us is that the players send a set of $\log n+O(1)$
qubits back and forth and apart from that no further qubits or
classical storage depending on the inputs are used. Also the protocol
runs in $O(\log n)$ stages, each concluded by a measurement. If this
measurement yields an index $i$ with $x_i=y_i=1$, then the protocol
stops (and rejects), else it continues. The qubits contain a superposition over
indices $i$ from $1$ to $n$ plus the values of $x_i$ and $x_i\wedge
y_i$. So an honest player that does not attempt to get more
information learns $O(\log n)$ times the measurement result for
$O(\log n)$ qubits and thus an information of at most $O(\log^2 n)$.

The main tool to show that the privacy loss is small against players
trying to get more information is the following
generalization of the famous no-cloning theorem \cite{WZ82}. While the
no-cloning theorem says that we cannot make a perfect copy of an
unknown quantum state (which would enable us to find out some
information about the state without changing the original by measuring
the copy), this lemma says that no transformation leaving two
nonorthogonal originals both
unchanged gives us any information about those states.

\begin{lemma}\label{lem:noinf}
Let $|\phi_1\rangle$ and $|\phi_2\rangle$ be two states that are
nonorthogonal. Assume a unitary map $U$ sends $|\phi_1\rangle\otimes |0\rangle$ to
$|\phi_1\rangle \otimes |a\rangle$ and $|\phi_2\rangle\otimes
|0\rangle $ to
$|\phi_2\rangle \otimes |b\rangle$. Then $|a\rangle=|b\rangle$.
\end{lemma}

{\sc Proof:}
The following simple proof has been proposed by Harumichi Nishimura [personal communication].

Note that the inner product of $|\phi_1\rangle$ and $|\phi_2\rangle$
is unchanged when we append some empty qubits, and when we apply the
same unitary operation to the states. Hence
\begin{eqnarray*}
\langle\phi_2|\phi_1\rangle&=&\langle\phi_2|\langle0|U^{\dagger}U|\phi_1\rangle|0\rangle\\
&=&\langle\phi_2|\langle b|\phi_1\rangle|a\rangle
               =\langle\phi_2|\phi_1\rangle\cdot\langle b|a\rangle.\end{eqnarray*}
Because $|\phi_1\rangle$ and $|\phi_2\rangle$ are nonorthogonal their
               inner product is nonzero and hence
we have $\langle b|a\rangle=1$. Therefore, $|a\rangle=|b\rangle$.
\qed

Now assume a protocol sends $k$ qubits (in a pure state) back and
forth without using
any private storage whose state depends on the input and without measuring (this
is what happens in the protocol for $DISJ_n$ during all $O(\log n)$ stages).
If we manage to change the messages in a way so that for no inputs
$x,x',y$ the message sent in round $t$ for input $x,y$ is orthogonal
to the message for input $x',y$, then there is no transformation for
Bob that
leaves the message unchanged, yet extracts some information. In other
words, honest players are forced to follow the protocol without
getting further information. The only information is revealed at the
end of a stage, when one player is left with the qubits from the last
message, resp.~at the time when one player decides to quit the
protocol. Thus at most ``size of the communication channel'' (i.e., $k$)
information is revealed.

We now describe how to make the messages nonorthogonal.

\begin{lemma}\label{lem:nonorth}
For all $\epsilon>0$ and
for any finite set of $l$-dimensional unit vectors $\{v_i\}$ there is a set of $l+1$
dimensional unit vectors $\{v_i'\}$  such that $||\tilde{v}_i-v_i'||<\epsilon$, and
$v_i'\bot v_j'$ for no $i,j$, where $\tilde{v}_i$ denotes $v_i$ with an appended 0 in dimension $l+1$.
\end{lemma}

{\sc Proof:}
For all vectors $v_i$ the vector $\tilde{v}_i$ is $l+1$ dimensional and contains the value 0 in the $l+1$st dimension
and the same values as $v_i$ in the other dimensions.
Then change $\tilde{v}_i(l+1)$ to $\delta$, and scale all other values by $\sqrt{1-\delta^2}$ to obtain $v'_i$. The
resulting vectors have norm $\langle
v'_i|v'_i\rangle=\delta^2+(1-\delta^2)\cdot\langle
v_i|v_i\rangle=1$. The inner product of two vectors is $\langle
v'_i|v'_j\rangle=\delta^2+(1-\delta^2)\langle v_i|v_j\rangle$. For a
finite set of, say $k$, vectors there are $k^2$ different values of
inner products $\langle v_i|v_j\rangle$. Using $\delta$ with
$-\delta^2/(1-\delta^2)$ different from all these values and small
enough that $||v_i-v_i'||\le\epsilon$ leads to a
set of vectors with the desired properties.
\qed

We can state the upper bound for disjointness.

\begin{theorem}\label{the:disjupper}
$DISJ_n$ can be computed by a quantum protocol with error $1/3$,
communication $O(\sqrt{n}\log n)$, and privacy loss $O(\log^2 n)$.
\end{theorem}

{\sc Proof:}
In \cite{BCW98} a quantum protocol with error $1/4$ and
communication $O(\sqrt{n}\log n)$ is described, in which Al and Bob
exchange pure state messages of length $\log n+O(1)$, but use no further storage
depending on the input. The protocol consists of $O(\log n)$ stages
each of which ends with a measurement of the qubits in the standard
basis. No further measurements are used.

We modify the protocol. We add one more qubit to the messages. Then we
change the first message to be sent (prepared by Al) as described in
lemma \ref{lem:nonorth}. The error introduced by this change is arbitrarily
small. Then the protocol is used as before, ignoring the new qubit in
all transformations, but always sending that qubit with the other
qubits back and forth between the players. This can be done in a way
ensuring that no message sent for any pair of
inputs $x,y$ in any round will be orthogonal to another such
message.

Assume Bob wants to get more information than he can get from the
$O(\log n)$ classical strings of length $\log n+O(1)$ obtained from the
measurements. In some round he will first start an attack on the
message. He has to map the message received to another message he must
send back. The second message is the result of a fixed unitary
transformation (depending on $y$) on the first. He has to combine the attack with
that unitary transformation. So we may assume that he first attacks
the message and then applies the transformation to get the next message.
The attack transformation maps the message and some empty qubits to the tensor
product of the same message and another state, that depends on the other
player's input. Lemma \ref{lem:noinf} ensures that this is
impossible. So Bob has to stick to the protocol without getting more
information than allowed.
\qed

Now we turn to the lower bound. Every classical deterministic protocol partitions
the communication matrix into rectangles labelled with the output of
the protocol. Let $\mu$ be a distribution on the inputs. A labelled rectangle
is $(1-\epsilon)$--{\it correct}, if according to $\mu$ at least $1-\epsilon$
of the weight of the rectangle is on correctly labelled inputs. Due to
Yao's lemma a randomized protocol with error $\epsilon$ and
communication $c$ yields for
every distribution $\mu$ a deterministic protocol that has error $\epsilon$
and the same communication. Such a protocol induces a partition of the
communication matrix into $2^c$ rectangles with overall error
$\epsilon$.

The {\it width} of a rectangle $A\times B$ is $\min\{|A|,|B|\}$. Let
$r(f)$ denote the largest width of any completely correct rectangle. \cite{BCKO93} proves:

\begin{fact}\label{fact:lower}
$CL_0(f)\ge (n-\log r(f))/2-1$ for all $f:\{0,1\}^n\times\{0,1\}^n\to\{0,1\}$.
\end{fact}

We now describe a new bound. An $a$-rectangle is a rectangle that contains predominantly the function value $a$.
The maximum size of a $(1-\epsilon)$--correct $a$-rectangle according to $\mu$ is
called $s^a_\epsilon(f,\mu)$. Let $uni$ denote the uniform distribution.

\begin{lemma}\label{lem:lower}
Choose $a\in\cal Z$. All randomized protocols with error $1/3$ computing
$f:\{0,1\}^n\times\{0,1\}^n\to\cal Z$ have privacy loss
\[\Omega\left(\frac{uni(f^{-1}(a))\cdot\log(1/s^a_{1/n^2}(f,uni))}{\log n}-O(1)
\right ).\]
\end{lemma}
{\sc Proof:}
Given a randomized protocol with error $1/3$ and privacy loss $c$ we
can find a protocol with privacy loss $c\cdot k$ and error
$1/2^{\Omega(k)}$ by repeating the protocol with independent coin
flips $k$ times and taking the majority output due to lemma \ref{lem:boost}.

For $k=O(\log n)$ we get a randomized protocol with privacy loss
$l/2=O(c\log n)$, and error $1/(2n^4)$. This privacy loss is guaranteed
against all distributions, and by one side of lemma \ref{lem:yao}
for the uniform distribution $uni$ there is a
deterministic protocol that has the privacy loss $l$,
and error $1/n^4$ on $uni$. The deterministic protocol
corresponds to a partition of the communication matrix into rectangles
with global error $1/n^4$, so $1-1/n^2$ of all inputs are in
rectangles that are $(1-1/n^2)$-correct.
Each $(1-1/n^2)$-correct $a$-rectangle has weight at most
$s=s^a_{1/n^2}(f,uni)$. Furthermore the total weight of inputs with function value $a$ in other rectangles is at most
$\alpha=1/n^4$.

After running the protocol we have a distribution on the values of Al's storage $A$ and
Bob's storage $B$. W.l.o.g.~both players have stored the complete
message history as a string $m$. Such a string is a label to a
rectangle in the communication matrix. Call such a rectangle $U_m\times V_m$ and
let $h(m)$ denote the height $|U_m|/2^n$, let $b(m)$ denote the base $|V_m|/2^n$.
$1-1/n^2$ of all inputs are in $(1-1/n^2)$-correct rectangles.
Let $M_a$ denote the set of $(1-1/n^2)$-correct
$a$-rectangles/message sequences in which the protocol outputs $a$.
$Pr(m)$ denotes the probability of rectangle $m$, i.e., its size under the uniform distribution.

The inputs that are not in $(1-1/n^2)$-correct $a$-rectangles but have function value $a$ have weight at most $\beta=1/n^4+1/n^2<2/n^2$.
They can contribute at most $\gamma=\beta n-\beta\log\beta=o(1)$ to an entropy
due to fact \ref{fact:maxent}.
Then\begin{eqnarray*}
&&H(Y|A,X,f(X,Y)=a)+H(X|B,Y,f(X,Y)=a)\\
&=& \sum_m Pr(m) [H(Y|m,X,f(X,Y)=a)+H(X|m,Y,f(X,Y)=a)]\\
&\le& \gamma  +\sum_{m\in M_a} Pr(m) [\log (2^n\cdot b(m))+\log(2^n\cdot h(m))]\\
&\le& o(1) + \sum_{m\in M_a} Pr(m) \log (|V_m|\cdot|U_m|)\\
&\le& o(1) + \sum_{m\in M_a} Pr(m) \log (2^{2n}s )   \\
&\le&2n+\log s+o(1).
\end{eqnarray*}
Also, assume that $uni^{-1}(a)\ge 1/n^2$, then
\begin{eqnarray*}
H(Y|X,f(X,Y)=a)\ge n-4\log n-o(1),\end{eqnarray*}
since in this case only $1/n^2$ of the weight of the distribution that is uniform on inputs with $f(x,y)=a$ can lie
on rows $x$ having
less than $2^n/n^4$ columns $y$ with $f(x,y)=a$. So if we pretend that all $x$ have at least $2^n/n^4$ corresponding
$y$ with $f(x,y)=a$, we increase the actual
entropy by at most $1/n^2\cdot n=o(1)$. But this would lead to $H(Y|X,f(X,Y)=a)\ge\log(2^n/n^4)$.
Consequently
\begin{eqnarray*}
uni(f^{-1}(a))\cdot H(Y|X,f(X,Y)=a)\ge uni(f^{-1}(a))\cdot (n-4\log n)-o(1).\end{eqnarray*}
\\This gives us
\begin{eqnarray*}
&&O(c\log n)\\
&\ge& I(Y:A|X,f(X,Y))+I(X:B|Y,f(X,Y))\\
&\ge& uni(f^{-1}(a))\cdot[ I(Y:A|X,f(X,Y)=a)+I(X:B|Y,f(X,Y)=a)]\\
&=&uni(f^{-1}(a))\cdot[ H(Y|X,f(X,Y)=a)+H(X|Y,f(X,Y)=a)\\
&&-H(Y|A,X,f(X,Y)=a)-H(X|B,Y,f(X,Y)=a)]\\
&\ge&uni(f^{-1}(a))\cdot[
2n-8\log n-2n-\log s] -  o(1)\\
&\ge&uni(f^{-1}(a))\cdot(
-\log (s))-O(\log n).\end{eqnarray*}

\qed

The following is proved in \cite{BFS86}
\begin{fact}\label{fact:Babai}
Let $\mu$ be the uniform distribution on pairs of sets of size
$\sqrt{n}$ from a size $n$ universe. Then the largest
$(1-\epsilon)$--correct 1-rectangle for disjointness  (i.e., one that contains mostly disjoint pairs of sets)
has size $1/2^{\Omega(\sqrt{n})}$ for
some constant $\epsilon$.
\end{fact}

\begin{corollary}\label{cor:disjlower}
$CL_{1/3}(DISJ_n)=\Omega(\sqrt{n}/\log n )$.
\end{corollary}

\section{The class of private functions}

We have seen in the previous section that certain functions can be
quantum  computed with less privacy loss against honest players  than possible
in the classical case. In this section we show that, however, the class
of functions which can be computed privately (i.e., with privacy loss
0) is unchanged by allowing quantum communication, if we consider
honest players (i.e., those who are not trusted to continue with the
protocol until the end).

\subsection{Players that do not preempt}
But first let us take a look at the model of honest players, in which
only the information retrievable at the end of the protocol is
counted.

\begin{theorem}
For every function $f$ with deterministic communication complexity $c$ there is a
quantum protocol with communication $O(c^2)$, where the
final state obtainable by every honest player has an arbitrarily small
distance to the final state of a player that knows only his input and
the function value at the end.
\end{theorem}

Thus we get an arbitrarily close approximation of privacy against
honest players if we consider only the information available at the
end of the protocol. In other words if we trust the other player not
only to play honest, but also to not quit before the end of the
protocol, every function can be computed in a secure way.

{\sc Proof:} Suppose there is a deterministic protocol for $f$ with
complexity $c$. First we turn this into a protocol, in which the
players do not need to store anything besides the current message,
i.e., they compute the new
message from the message they received, send the message, and remember
nothing else. For this the players simply exchange a complete message
history in all the rounds, increasing the complexity to $c^2$ at
most.
Now following lemma \ref{lem:nonorth} we can turn this into a quantum
protocol with arbitrary small error $\epsilon$ and communication
$c^2+c$, in which only pure state messages are exchanged, so that for no
inputs $x,y,y'$ the messages on $x,y$ and on $x,y'$ sent to Al (or
Bob) in some
round $t$ are orthogonal. With lemma \ref{lem:noinf} then an honest Al
cannot obtain information from the message he holds without changing
the message similarly to the proof of theorem \ref{the:disjupper},
thus Al has to send the message and is left with no
information in all rounds after the message is sent.

At the end, however a complete message history is available to one
player, making the protocol highly nonprivate. To remove this problem
consider the following. A {\it clean} protocol
\cite{CDNT98} is a protocol, in which the final state is \[|0\rangle
|x\rangle|f(x,y\rangle|y\rangle|0\rangle.\] \cite{CDNT98} shows how to
transform any quantum protocol with error $\epsilon$ into a protocol,
whose final state has distance $O(\sqrt{\epsilon})$ from the final
state a clean protocol would have. We use this transformation, which
also increases the communication complexity by a factor of 2 only and
does not change the error (the idea is that the ``garbage'' produced by the
computation is removed by ``reversing'' the protocol).

Thus we get a protocol with error $O(\sqrt{\epsilon})$, communication $O(c^2)$,
in which in all rounds the players exchange a certain set of qubits,
about which both players cannot obtain additional information, since these
messages are pairwise nonorthogonal. In the end the state has arbitrarily small
nonzero distance to a state revealing no additional information.
\qed

Due to the continuity of entropy described by fact~\ref{fact:contin} both the
(information-theoretically measured) privacy loss and the
(distance measured) leakage can be
made arbitrarily small at the end of a protocol.

\subsection{The characterization of quantum privacy}
Now we return to our regular definition of privacy and show that here
quantum communication does not enlarge the set of private functions.
The set of classically private functions has been characterized in
\cite{K92} and \cite{B89}. We extend this characterization to the
quantum case.

\begin{definition}
Let $M=C\times D$ be a matrix. A relation $\equiv$ is defined as
follows: rows $x$ and $x'$ satisfy $x\equiv' x'$, if there is a column
$y$ with $M_{x,y}=M_{x',y}$. Then $\equiv$ is the transitive closure of
$\equiv'$. Similar relations are defined for columns.

A matrix is called forbidden, if all its rows are equivalent, all
its columns are equivalent, and the matrix is not monochromatic.
\end{definition}

\begin{theorem}\label{the:char}
If the communication matrix of $f$ contains a forbidden submatrix then
$f$ cannot be computed by a quantum protocol with error
smaller than 1/2 and no privacy loss.
\end{theorem}

{\sc Proof:}
A quantum protocol with error smaller than $1/2$ for $f$ must also solve
the problem $g$ corresponding to the forbidden submatrix. If $A$
contains the row-labels and $B$ the column labels of the forbidden
submatrix then $g$ is defined on $A\times B$ and $g(x,y)=f(x,y)$.
This problem $g$ is
nontrivial, since the submatrix is not monochromatic. Suppose a given
protocol computes $g$ with error smaller than $1/2$ and privately. We will show
that one round after the other can be shaved off the protocol,
eventually yielding a protocol for $g$ with one round. Such a protocol
cannot compute $g$ with error smaller than 1/2, thus we reach a
contradiction.

We show that the first message (w.l.o.g.~sent by Al) does not depend
on the input, and can thus be computed by Bob, whereupon the first
round of communication can be skipped. Let $x_1,\ldots,x_l$ denote the
rows of the forbidden submatrix, enumerated in such a way that
$x_i\equiv' x_j$ for some $j<i$ for all $i>1$. If $x_i\equiv' x_j$
then there is a $y$, so that $g(x_i,y)=g(x_j,y)$. Since it is possible
that Bob holds $y$, Al is not allowed to send different
messages on $x_i$ and $x_j$, since otherwise Bob may obtain
information about the identity of Al's inputs $x_i$ and $x_j$ not deducible from the function
value alone. So for all $x_i$ the same message is sent.

Let $\rho^{xy}_{AM}$ denote the state of Al's qubits right before the first
message is sent (on inputs $x,y$), with $M$ containing the message.
$\rho^{xy}_M$ is the same for all $x,y$. Also $\rho_{AM}^{xy}$
purifies such a state. Due to the local
transition theorem (fact \ref{fact:local}) Al has unitary operations acting
on register $A$ that switch  between those states (for different $x$) without introducing any error. Thus
Bob may prepare $\rho_{AM}^{x,y}$ for some fixed $x$, send the part
of the state in $A$ to Al and keep the $M$ part. Al can then change
the received state to the one for the correct $x$. Furthermore Bob can
send the message for round 2 together with the first message, thus we
get a protocol with one round less.

Repeating this we eventually arrive at a protocol with one round only, in which, say,
a message is sent from Al to Bob. Thus if the error is smaller than
1/2 the output does not depend on
$y$. Consequently the communication matrix of $g$ consists of monochromatic
rows only, and there are at least two different such rows, since the
matrix is not monochromatic. Such a matrix is clearly not a forbidden submatrix, since two
different monochromatic rows are not equivalent. Thus we arrive at a
contradiction to our assumptions on $f$ or on the protocol.  \qed

Now that we know a forbidden submatrix excludes a private quantum
protocol, the other piece for a characterization is as follows, see
\cite{K92}.

\begin{fact}
If the communication matrix of $f$ contains no forbidden submatrix,
then $f$ can be computed by a deterministic private protocol.
\end{fact}

Thus the class of privately computable functions is invariant under
the choice of quantum or classical communication.

A function $f$ can be computed with privacy loss $k$ in the hint sense, if
there is a privately computable function $h$, such that $f(x,y)$ can be computed from
$h(x,y)$, and $k=\log(range(h))-\log(range(f))$. Since a function $h$
can be computed privately deterministically, iff $h$ can be computed
privately by a quantum protocol, we get the following.

\begin{corollary}\label{cor:hint}
 The privacy loss of a function $f$ in the hint
  sense is unchanged if we allow quantum protocols.\end{corollary}

The structure imposed on protocols by the privacy constraint is
actually strong enough to deduce a lower bound on the number of rounds
needed to compute a function.

\begin{theorem}\label{the:rounds}
Any function $f$ computable by a private quantum protocol with error
smaller than 1/2 and $r$ rounds of communication can also be computed
by a private deterministic protocol with no error using at most $r$ rounds.
\end{theorem}

{\sc Proof:}
We construct a {\it protocol tree} from the quantum protocol. This is
a layered directed tree whose vertices are indexed with rectangles in the communication
matrix. Rectangles $A\times B$ in depth $d$ have children $A_i\times
B$ with disjoint $A_i$ covering $A$, or children $A\times B_i$ with
disjoint $B_i$ covering $B$. In depth $d$ either all edges lead to
vertices that decompose the set of rows or all edges lead to vertices that
decompose the set of columns.

The root is indexed by the communication matrix $M=A\times B$ of $f$. W.l.o.g.~assume
Al sends a message in the first round. Then the set of messages used
by Al decomposes the set of rows into disjoint subsets. Note that if
$x\equiv x'$ for two inputs $x,x'$ to Al then these inputs share the
same message in the first round. Recall that a message is in general a
mixed quantum state. If Al's messages induce subsets
$A_1,\ldots,A_t$ of the rows, and the equivalence relation $\equiv$ on
rows (relative to $M$) has equivalence classes $C_1,\ldots,C_l$ then each $C_i\subseteq
A_j$ for exactly one $j$.

Now from the point of view of Bob all rows in some set $A_i$ are
equivalent when he sends his message in round 2. Hence we may as
before decompose the columns of each rectangle $A_i\times B$ according
to the messages used by Bob. Again any equivalence class $C_i$ for columns
(where the equivalence relation $\equiv$ is chosen relative to
$A_i\times B$) lies in exactly one subset $B_j$ of the row
decomposition induced by Bob's messages.

In this manner we can inductively follow the protocol round per round
 to find a protocol
  tree. Note that all inputs in the rectangles attached to the leaves
  of the tree have the same acceptance probabilities, which are either
  all smaller than 1/2, or all larger than 1/2. Hence it is true that
  the rectangles attached to leaves are monochromatic.

A protocol tree trivially induces a deterministic protocol for $f$ using as
many rounds as the tree is deep, while the depth of the constructed
tree is the number of rounds of the quantum protocol.
\qed

Since the deterministic protocol constructed for the previous theorem
trivially doesn't have to communicate more than $n$ bits in one round, its
communication cost is at most $n$ times higher than the number of
rounds of the quantum protocol, which is a lower bound on the quantum
communication cost.

\begin{corollary}\label{cor:comm cost}
The communication cost of an optimal private quantum protocol for a
function $f:\{0,1\}^n\times\{0,1\}^n\to\N$ is at most a factor $n$
smaller than the communication cost of an optimal private
deterministic protocol for $f$.
\end{corollary}

\subsection{Boolean functions and leakage}
Next we consider
the case of Boolean functions. It is known \cite{CK91} that the class of
private Boolean functions is the class of functions $f_A(x)\oplus
f_B(y)$, even if one considers protocols that leak $\delta$ (recall
that this refers to the
distance sense of leaking) and have
error $\epsilon$ with $\epsilon+\delta<1/2$. These functions are
combinatorially characterized by the so-called ``corners lemma''
\cite{CK91} saying that there is no $2\times 2$ rectangle in the
communication matrix containing 3 ones and 1 zero or vice versa.
As a corollary of theorem \ref{the:char} we get a result for the quantum
case with no leakage.

\begin{corollary}\label{cor:corners}
If the communication matrix of a function $f$ contains a $2\times 2$
rectangle with exactly 3 times the same entry, then no private quantum protocol with
error smaller than 1/2 can compute $f$.
\end{corollary}

\begin{corollary}\label{cor:XOR}
The class of Boolean functions computable by private quantum protocols
is the class of functions $f_A(x)\oplus f_B(y)$.
\end{corollary}

Is corollary \ref{cor:corners} also valid in the quantum case with
small leakage? The
answer is no. There are function satisfying the assumptions of the
corners lemma which
can be computed with small leakage by a quantum protocol.

\begin{theorem}\label{the:AND}
There is a quantum protocol computing the AND function on two bits
with error 1/3 that has leakage $\delta$ and uses $O(1/\delta^2)$ communication.
\end{theorem}

{\sc Proof:}
We describe a protocol in which only $\delta$ is leaked to Bob, and
nothing is leaked to Al. During $O(1/\delta^2)$ rounds Al prepares a
superposition
$\delta/2 |00\rangle + \sqrt{1-\delta^2/4}|11\rangle$ if $x=0$,
and $\delta/2 |10\rangle+ \sqrt{1-\delta^2/4}|11\rangle$ if $x=1$.
Note that the trace distance between the corresponding density
matrices is $\delta$ due to equation (1). Thus if Bob receives such a message leakage to him is $\delta$.
Bob then adds a blank qubit and if $y=1$ applies a
unitary transformation that sends $|000\rangle$ to $|000\rangle$,
$|100\rangle$ to $|101\rangle$, and $|110\rangle$ to $|110\rangle$. If $y=0$ he leaves the state unchanged. Then
Bob sends the 3 qubits back to Al. Al and Bob repeat this
$O(1/\delta^2)$ times with fresh qubits. In the end Al measures all triples in the
standard basis. If he receives a $|000\rangle$ state he outputs 0,
 if he gets a $|101\rangle$ state he outputs 1 (and sends the result
to Bob). If he has no such results he gives up without answer.

Note that with probability $\delta^2/4$ Al gets one of the desired
results, thus $O(1/\delta^2)$ experiments suffice to yield a protocol
with constant error.

The leakage can be analyzed as follows. Suppose $x=y=1$. In this case
there can be no leakage, since an input and the function value give away
the other input.

Now suppose $x=1,y=0$. In this case there can be no leakage to
Al. The leakage to Bob is $\delta$ given the message of a
round. Due to lemma \ref{lem:noinf} Bob cannot get any information out of a message
without becoming dishonest, since no two messages are
mutually orthogonal. Thus for all rounds the information leaked to Bob is
$\delta$.

Suppose $x=0$. Al always simply gets his message back, no matter what
Bob's input is, so there is no leakage to Al. If $y=1$ then there can
be no leakage to Bob. Otherwise the leakage to Bob
is $\delta$ as above.
\qed

The communication complexity in the above construction is within a
polynomial of the optimum.

\begin{theorem}\label{the:cornersleak}
If the communication matrix of a function $f$ contains a $2\times 2$
rectangle with exactly 3 times the same entry, then no  quantum protocol with
error $1/3$, leakage $\delta$, and at most $1/(12\sqrt{\delta})$ rounds can
compute $f$.
\end{theorem}

{\sc Proof:}
A protocol containing the described submatrix can easily be adapted to
compute the Boolean AND function on input bits $x,y$ with the same parameters.
We show that the above stated number of rounds is necessary.

Let $\rho_{AM}^{x}$ denote the state of the private qubits of Al in
$A$ and of the first message sent to Bob in $M$ in round one
when Al's input is $x$. Then the given leakage demands that
$\trn{\rho_M^{1}-\rho_M^{0}}\le\delta$, since it is possible that
Bob's input is $y=0$.
The states $\rho_{AM}^{1}$ and $\rho_{AM}^{0}$ are purifications of the two
states on the message qubits.

Due to the local
transition theorem (fact \ref{fact:local})
there is a unitary transformation $U$ on Al's qubits
alone that maps $\rho_{AM}^0$ to a state $\tilde{\rho}_{AM}^{0}$ with
\[\trn{\tilde{\rho}_{AM}^{0}-\rho_{AM}^{1}}\le 2\sqrt{\delta}.\]

We modify the protocol by skipping the first round. Instead Bob
creates a state $\rho_{AM}^0$ by himself and sends the $A$ part to Al
together with the communication of round 2. If Al's input is 0, the
protocol can continue without problems. If Al's input is 1, he applies
the unitary transformation $U$, which leads to the state
$\tilde{\rho}_{AM}^0$ with distance $2\sqrt{\delta} $ from $\rho_{AM}^1$.
So the error introduced is at most $2\sqrt{\delta}$ and the protocol
runs with one round less.

Repeating the above process for a $k$ round protocol leads eventually
to a protocol in which, say, Al sends one message and Bob none, so the
output does not depend on Bob's input anymore. This can only happen when
the error is at least 1/2, so $1/3+(k-1)\cdot 2\sqrt{\delta}\ge1/2$, hence
$k>1/(12\sqrt{\delta})$.
\qed

\section{Trading privacy loss against complexity}

In this section we show that allowing a privacy loss of much less than one
bit (instead of privacy loss 0) can reduce the communication complexity of a
certain function, namely the identified minimum problem,
from exponential to polynomial. Thus protocols obtaining a very close approximation of privacy can be much
cheaper than truly  private protocols.

\begin{theorem}\label{the:save}
The function $IdMin_n$ can be computed by a randomized protocol with
privacy loss $\delta$, error $\delta$, and
communication $O(n^3/\delta\cdot \log(1/\delta) )$.

Every quantum or randomized protocol computing $IdMin_n$ with error
$\epsilon<1/2$ and with privacy loss 0 needs communication
$\Omega(2^n)$.
\end{theorem}

{\sc Proof:} It is shown in \cite{K92} that any private randomized or
deterministic protocol for $IdMin_n$ needs $2\cdot(2^n-1)$
communication rounds. With theorem~\ref{the:rounds} this implies that
also quantum protocols need that many rounds. The communication cost
is always at least as large as the number of rounds.

For the upper bound we proceed as follows. We first show that the
function can be computed efficiently with small leakage. Then we invoke lemma
\ref{lem:leakpriv} to get the result for privacy loss. To
construct a protocol with small leakage
we describe for every probability distribution on the inputs a
deterministic protocol that has small expected leakage (over the input
distribution). Using Yao's lemma like in lemma \ref{lem:yao} we get a single
randomized protocol that has small expected leakage against all
distributions, where the expectation is over the coins of the
protocol. Such a protocol immediately has small leakage in the sense
of our standard
definition. To remove the necessity of allowing public coin randomness we
use lemma \ref{lem:Newm}. Then we are ready to guarantee also small
privacy loss using lemma \ref{lem:leakpriv}.

The expected leakage to Al in a protocol for a distribution $\mu$ is
\[E_{x,y,y':f(x,y)=f(x,y')} ||\rho_A^{xy}-\rho_A^{xy'}||_1,\] for the
state $\rho^{xy}_{AB}$ of the storage of Al and Bob in some
round $t$. The expected leakage of the protocol is the maximum over
all rounds and over Al,Bob.

The corresponding Yao lemma is as follows, being proved completely analogous to lemma~\ref{lem:yao}.
Note that if a randomized protocol has
for all inputs an expected leakage of at most $\zeta$ (with the expectation
over its coins), then it has leakage $\zeta$ in the ordinary
definition.

\begin{lemma}\label{lem:Yao2}
The following statements are equivalent in the sense that if one is true for some values $\epsilon,\zeta$, then
the other is true with values $2\epsilon,2\zeta$.
\begin{itemize}
\item There is a randomized public coin protocol for a function $f$ with
communication $c$,
error $\epsilon$, and leakage $\zeta$.
\item For every distribution $\mu$ on inputs there is a deterministic
protocol for $f$ with communication $c$, and error $\epsilon$ and
expected leakage $\zeta$ on $\mu$.
\end{itemize}\end{lemma}

We start by describing a protocol with small expected leakage
for the uniform distribution and then show how to adapt this protocol
to an arbitrary distribution.

The protocol is defined inductively. For $n=O(1)$ we use the simple
protocol with leakage 0, in which Al asks for $z=1,\ldots,x-1$, whether
$z\ge y$. If so for one $z$, then $2y$ is the result, otherwise $2x+1$ is the
result. The protocol needs communication $O(1)$.

For larger $n$ we do the following.
Let $\gamma=\delta/(16n) $. Al asks Bob for all $z_i=\lceil (1+\gamma)^i
\rceil<\min\{x,2^{n-1}\}$ (with $i\in\N)$, whether $z_i\ge y$. If so for one $z_i$, then $2y$ is the
output (given by Bob). Else, when $x\le 2^{n-1}$, then
$2x+1$ is given as output. If this is not the case, then both players know
the minimum is larger than $2^{n-1}$, and the
protocol can be invoked recursively for $n-1$.

First we compute the communication cost of the protocol.
Obviously the communication before the recursion is in
$\log_{1+\gamma} 2^n=O(n/\gamma)=O(n^2/\delta )$ rounds, each
communicating at most $n$ bits. So the recursion for the overall
communication is $C(n)=O(n^3/\delta )+C(n-1)$. After $\log(1/\delta)$ recursions, however, the remaining pairs
of inputs have weight at most $\delta^2$. So we may stop there, and hence the communication is at most
$O(n^3/\delta\cdot \log(1/\delta))$.

Next we compute the leakage and error of the protocol. Let $\alpha=(1+\gamma)^i\le x\le
(1+\gamma)^{i+1}<2^{n-1}$. Also assume that we are in an iteration,
where the remaining input length is $n$ (which is not necessarily
the original input length). If the protocol stops early, the error is at most $\delta^2$, and the leakage is 0. Otherwise
the following holds.

If $y\ge x$, then the protocol will neither err nor leak information.

If $\alpha\le y< x$, then the
protocol will err and will leak, since Bob learns $x$. This happens with probability at most
\[\left((1+\gamma)^{i+1}-(1+\gamma)^i\right)/2^n=
\gamma (1+\gamma)^i/2^n\le \gamma/2,\] and the leakage contributed by this case is at most
$2\cdot \gamma/2=\delta/(16n)$, the error at most $\gamma/2$.

If $y<\alpha$, then the protocol is correct. There is, however, some
leakage. Bob learns that $x>(1+\gamma)^i$ for some $i$, instead of
just learning that $x>y$. This corresponds to Bob knowing that $x$ is
distributed uniformly over all values larger than $(1+\gamma)^i$ instead
of Bob knowing that $x$ is distributed uniformly over all values larger than
$y$. The first distribution $p$ is uniform on
$2^n-(1+\gamma)^i=2^n-(1+\gamma)^{i-1}-\gamma(1+\gamma)^{i-1}$ values,
the second distribution $q$ is uniform on $2^n-y$ values.
The distance between $p$ and $q$ is at most the distance
between $p$ and the distribution $q'$ in which $x$ is
uniform over all values larger than $(1+\gamma)^{i-1}$.
Then $q'$ is uniform on $2^n-(1+\gamma)^{i-1}$ points.
Since $(1+\gamma)^{i-1}\le 2^n/2$, the distance is at most $2\gamma$
and the probability of this event is at most 1/2, so the leakage contributed by this case is at most $\gamma= \delta/(16n).$

Thus the overall leakage is smaller than $\delta/(8 n)$, and the error
$\gamma/2+\delta^2<\delta/4$.

After describing the protocol for the uniform distribution we now turn
to protocols for arbitrary distributions $\mu$. Since Al plays the
role of an interrogator in the above protocol, while Bob only answers,
let $x$ be an arbitrary input for Al and $\mu_x$ the induced distribution on Bob's
inputs. Note that this distribution is known to Al.

Let $r_l$ be the least integer satisfying $\sum_{i=1}^{r_l}\mu_x(i)\ge
l/2^n$ for $l=1,\ldots,2^n$. The protocol proceeds as in the protocol
for the uniform distribution, but Al queries $r_l$ instead of $l$ all
the time, i.e., for $l=(1+\gamma)^i<\min\{x,r_{2^{n-1}}
\}$.

The communication complexity of the protocol is still $O(n^3/\delta\cdot\log(1/\delta))$.

Let $\alpha=r_{(1+\gamma)^i}\le x\le r_{(1+\gamma)^{i+1}}$.

If $y\ge x$, then the protocol will
neither err nor leak information.

If $\alpha< y< x$, then the
protocol will err and will leak information.
The probability that this happens is at most
$(1+\gamma)^{i+1}/2^n-(1+\gamma)^i/2^n\le \gamma\frac{(1+\gamma)^i}{2^n}
\le\gamma/2$, and the leakage in this case is at most
$\gamma$, the error at most $\gamma/2$.

If $y<\alpha$, then the protocol is correct, but again there is leakage.
Bob gets to know that $x\ge\alpha$ instead of just knowing $x>y$, i.e.,
the (normalized) distribution on values larger than $y$ against
the (normalized) distribution on values larger $\alpha$. The distance
between these two distributions is at most $2\gamma$. Thus the
contribution to the expected leakage is at most $\gamma.$

The expected leakage of all cases together is at most $\delta/(8n)$.

So we get for every distribution a deterministic protocol with error
$\delta/4$ and expected leakage $\delta/(8n)$. Then lemma \ref{lem:Yao2} gives us a
single public coin randomized protocol with leakage $\delta/(4n)$.
The communication complexity is $O(n^4/\delta)$, the error is at most
$\delta/2$.

Applying lemma \ref{lem:Newm} gives us a protocol with
leakage $\delta/(2n)$, error $\delta$, communication $O(n^4/\delta)$
using no public coin.
Then we can use lemma \ref{lem:leakpriv}
to see that the protocol actually has privacy loss at most $\delta$.\qed

\section{Conclusions and open problems}

In this paper we have discussed privacy with respect to honest players
with a focus on the themes quantum communication and protocols with (small)
privacy loss or leakage. We have given an example of a function that
can be computed with exponentially smaller privacy loss using quantum
communication than in the case of classical communication. The set
of functions with privacy loss 0 is, however, not enlarged by quantum
communication. For Boolean functions we were able to give a simple
characterization of the quantum private functions as $f_A(x)\oplus
f_B(y)$. It is known \cite{CK91} that allowing small leakage (leakage
$\delta$ and error $\epsilon$ with $\epsilon+\delta<1/2$) for classical
communication does not allow to compute more functions. In the quantum
case, however, leakage allows to compute the AND function (with a
tradeoff between the number of rounds and the leakage).

The characterization for Boolean functions can be extended to the case
of multiparty private computation. As in \cite{CK91} it can be shown
that only functions of the form $f_1(x_1)\oplus\cdots\oplus f_k(x_k)$
can be quantum computed in a way so that every set of $\lceil
k/2\rceil$
players learns nothing more about the other players' inputs than what
is deducible from
their inputs and the output alone. Since every function can be
computed classically so that no coalition of less than $k/2$ players
learns more than allowed \cite{BGW88,CCD88}, and the aforementioned
functions are private against coalitions of even $k-1$ players, there
are only 2 levels in this hierarchy of privacy for quantum computation,
as in the classical case, see \cite{CK91}. Note that more such levels
exist for non-Boolean functions \cite{CGK94}.

We now give some open problems. A more realistic type of player is
an adversary that has two objectives: with highest priority he wants
the output to be correct with large probability. But then he also wants
to learn as much as possible under this constraint.
As an illustration of the power of this kind of player consider
a technique from a proof for a lower bound on the quantum communication complexity of
the inner product function in \cite{CDNT98}. Given any clean
(for simplicity assume errorless) protocol
for the inner product function, one player may take a uniform
superposition over all possible inputs instead of his real input and
use that protocol. Applying a Hadamard transform to his (fake) input register
after the protocol has stopped supplies the player with the other
player's input. So he is able to compute the function value and learn
maximal information at the same time given any clean protocol. Note
that the player is not even approximately honest, though.

A restricted form of this hard to analyze type of player is an
{\it almost honest} player that roughly follows the protocol,
but only sends messages that are in distance $\epsilon$ from the
``correct'' messages. This allows in the quantum case
e.g.~to use approximate cloning as in \cite{BH96}, or generally the following
type of attack: the player uses the protocol with some probability
$\epsilon$ for a fake input. Then he learns some information he should
not know with probability $\epsilon$. If we measure the divulged
information in the information theoretic sense there are private
functions, e.g.~$IdMin_n$, for which such a player can obtain
$\epsilon n$ bits of information while being approximately honest.

A study of privacy with approximately honest players would be
 interesting. In particular, can the
quantum protocol for disjointness be made secure against them?
The set of {\it private} Boolean functions is robust
against such players, since they are of the form $f_A(x)\oplus
f_B(y)$. About non-Boolean functions no results seem to be known even in the
classical case.

Another open problem is the following: can we extend the
characterization of (non-Boolean) classically private
functions for the case of small leakage, or does small leakage make
some (non-Boolean) nonprivate functions computable in the classical case?

Finally, how can one prove lower bounds on the privacy loss of quantum protocols?
Since the privacy loss is always smaller than the communication
complexity, this is different from proving lower bounds for quantum communication complexity
as e.g.~in \cite{K01,R03}. We have shown that the
quantum privacy loss of the disjointness problem is only $O(\log^2 n)$, while the
quantum communication complexity of disjointness is
$\Omega(\sqrt{n})$ \cite{R03}.

\section*{Acknowledgements}
The author wishes to thank Harumichi Nishimura for useful comments
improving the presentation of lemma 12, Hoi-Kwong Lo for
interesting discussions, and the referees for useful comments.

\end{document}